\RequirePackage{etoolbox}
\let\originalshowhyphens\showhyphens

\documentclass[sigconf]{acmart}

\let\showhyphens\originalshowhyphens

\usepackage{tcolorbox}
\usepackage{multirow}
\usepackage{siunitx}
\usepackage{arydshln}
\usepackage[table]{xcolor}
\usepackage{booktabs}
\usepackage{url}
\usepackage{graphicx}

\tcbset{
    mysummarybox/.style={
        colback=blue!5!white,
        boxrule=0pt,
        colframe=blue!5!white,
        sharp corners,
        boxsep=5pt,
        left=0pt, right=0pt, top=0pt, bottom=0pt,
        fontupper=\normalsize
    }
}

\usepackage{enumitem}
\newlist{rquestions}{enumerate}{1}
\setlist[rquestions]{
    label=\textbf{RQ\arabic*:},
    labelwidth=*,
    font=\itshape,
    align=left
}

\copyrightyear{2026}
\acmYear{2026}
\acmConference[MSR '26]{23rd International Conference on Mining Software Repositories}{April 13--14, 2026}{Rio de Janeiro, Brazil}
\acmPrice{}
\acmDOI{}

\begin{document}

\title{When AI Agents Touch CI/CD Configurations: Frequency and Success}

\author{Taher A. Ghaleb}
\orcid{0000-0001-9336-7298}
\affiliation{
  \department{Department of Computer Science} 
  \institution{Trent University}
  \city{Peterborough}
  \state{Ontario}
  \country{Canada}}
\email{taherghaleb@trentu.ca}

\begin{abstract}
AI agents are increasingly used in software development, yet their interaction with CI/CD configurations is not well studied. We analyze 8,031 agentic pull requests (PRs) from 1,605 GitHub repositories where AI agents touch YAML configurations.
CI/CD configuration files account for 3.25\% of agent changes, varying by agent (Devin: 4.83\%, Codex: 2.01\%, $p<0.001$). When agents modify CI/CD, 96.77\% target \textsf{GitHub~Actions}. Agentic PRs with CI/CD changes merge slightly less often than others (67.77\% vs.\ 71.80\%), except for Copilot, whose CI/CD changes merge 15.63~percentage points more often. Across 99{,}930 workflow runs, build success rates are comparable for CI/CD and non-CI/CD changes (75.59\% vs.\ 74.87\%), though three agents show significantly higher success when modifying CI/CD.
These results show that AI agents rarely modify CI/CD and focus mostly on \textsf{GitHub~Actions}, yet their configuration changes are as reliable as regular code. Copilot’s strong CI/CD performance despite lower acceptance suggests emerging configuration specialization, with implications for agent training and DevOps automation.
\end{abstract}



\maketitle

\section{Introduction}
AI agents are increasingly integrated into software development workflows, generating code at scale~\cite{li2025aidev,watanabe2025use,perry2023users}. Prior work evaluates their code generation, bug fixing, and security~\cite{nguyen2022empirical,sandoval2023lost,asare2023github}, but focuses primarily on application code. Yet, modern software development relies heavily on CI/CD pipelines, which automate testing and deployment via declarative YAML configurations that directly affect software reliability~\cite{hilton2016usage,zhao2017impact,shahin2017continuous,fitzgerald2017continuous}. Whether AI agents can be trusted to modify CI/CD configurations remains unknown~\cite{ghaleb2025llm4ci}.

In this paper, we investigate AI agent interactions with YAML configurations in 8,031 agentic pull requests (PRs) from 1,605 GitHub repositories, covering five agents (Copilot, Cursor, Devin, Claude Code, and OpenAI Codex). We observe that AI agents change YAML files in only 3.25\% of changes, with significant variation across agents ($\chi^2=3016$, $p<0.001$). Also, when agents change CI/CD, \textsf{GitHub~Actions} dominates (96.77\%), indicating strong platform concentration. Besides, our results reveal that PRs with CI/CD changes are merged slightly less often than others (67.77\% vs.\ 71.80\%), except for Copilot, which shows higher acceptance for CI/CD changes of +15.63~percentage points (pp). Analysis of 99,930 workflow executions shows comparable build success for CI/CD and non-CI/CD changes (75.59\% vs.\ 74.87\%, $p=0.138$), with three agents achieving statistically significant improvements when changing CI/CD.

\smallskip
\noindent\textbf{Contributions:}
(1) We empirically study AI agents’ interactions with CI/CD configuration across 711,923 file changes;  
(2) We provide the first analysis of platform concentration and agent specialization in CI/CD, highlighting Copilot’s distinct behavior;  
(3) We present evidence on the reliability of AI-generated CI/CD changes, informing trust and design considerations for AI-assisted DevOps.

\smallskip
\noindent\textbf{Paper organization:}
Section~\ref{sec:background} gives some background about AI agents and CI/CD.
Section~\ref{sec:empirical} presents our data source and three research questions (RQs) analysis and results. 
Section~\ref{sec:implications} discusses implications. 
Section~\ref{sec:threats} discusses validity threats. 
Section~\ref{sec:related} reviews related work. 
Section~\ref{sec:conclusion} concludes the paper.

\smallskip
\noindent\textbf{Replication:} Our data and scripts available at~\cite{replication_package}.

\section{Background}
\label{sec:background}
\noindent\textbf{AI Agents.} We examine five prominent AI agents: \textsf{GitHub Copilot} provides real-time IDE-integrated code suggestions; \textsf{Cursor} offers an AI-first code editor with deep codebase understanding; \textsf{Devin} acts as a fully autonomous AI software engineer; \textsf{Claude Code} provides strong reasoning; and \textsf{OpenAI Codex} powers various coding interfaces. Their differing architectures and training methods may yield distinct behaviors towards configurations.

\smallskip
\noindent\textbf{Continuous Integration and Delivery (CI/CD).} CI/CD automates code verification through building and testing, while CD extends this to production~\cite{fowler2024continuous}. Modern CI/CD platforms (e.g., \textsf{\textsf{GitHub~Actions}}, \textsf{Travis~CI}, \textsf{CircleCI}, and \textsf{GitLab~CI}) automate workflows through declarative \textsf{YAML} configuration files defining triggers, jobs, steps, environments, and conditions that directly impact software reliability~\cite{zhao2017impact}. These pipelines detect integration issues early, enforce consistency, and improve maintainability.

\vspace{-2pt}
\section{Empirical Analysis and Results}
\label{sec:empirical}

\subsection{Data Source}
We use the AIDev-pop dataset~\cite{li2025aidev} (Oct 28, 2025 update), containing 33,596 PRs created by five AI agents across GitHub repositories. We processed 711,923 file changes from 8,031 PRs in 1,605 repositories where AI agents modified YAML files. Using GitHub's GraphQL API, we augmented the dataset with \textsf{GitHub~Actions} workflow execution data, collecting 99,976 workflow runs. We constructed a list of 24 CI/CD services with regular-expression-based detection patterns, validated on 23,161 YAML changes for statistical analysis.

\vspace{-2pt}
\subsection{RQ1: How do AI agents interact with CI/CD services and their configurations?}

\subsubsection{\textbf{Motivation.}}
Studying how AI agents interact with CI/CD configurations can reveal whether they treat them as a core or peripheral concern. CI/CD YAML files define automation workflows that directly affect software reliability~\cite{ghaleb2019empirical,zheng2025github,ghaleb2019noise,ghaleb2022interplay}. If agents change these configurations much less often than application code, this may indicate gaps in their training data or capabilities. Moreover, identifying which CI/CD platforms agents target more can reveal training biases and potential lock-in.

\subsubsection{\textbf{Approach.}}
We analyzed 711,923 file changes, identifying YAML files by extension and tracking changes at file, PR, commit, and project levels. We performed chi-square tests~\cite{mchugh2013chi} to assess whether agents differ significantly in YAML modification rates. For CI/CD platform detection, we developed regular expression patterns (see our replication package~\cite{replication_package}) for 24 CI/CD services and classified YAML changes to identify CI-specific changes.

\medskip
\subsubsection{\textbf{Findings.}}
~

\smallskip
\noindent\textbf{Agents change YAML in only 3.25\% of file changes, with significant agent variation.}
Our analysis revealed that AI agents modified YAML files in only 23,161 of 711,923 file changes (3.25\%), indicating these artifacts receive far less attention than application code. Table~\ref{tab:agent_yaml_rates} shows substantial variation across agents: Devin has the highest rate (4.83\%), OpenAI Codex the lowest (2.01\%), a 2.4$x$ difference confirmed as statistically significant ($\chi^2=3016.26$, $p<0.001$).
The distribution is highly right-skewed: the mean of 17.63 versus the median of 3.0 YAML file changes per project reflects that 25\% of projects have only one file change and 75\% have fewer than 9, yet the maximum reaches 3,589. At the PR level (mean=5.0, median=2.0), 47\% of projects have exactly one PR touching YAML, a ``one-and-done'' pattern suggesting agents treat CI/CD as auxiliary configuration. Most projects change a single YAML file (median=1.0), though some touch hundreds (max=810).
Concentration analysis indicates systematic configuration management: the top 10 projects account for 33.68\% of YAML file changes but only 12.61\% of PRs and 12.12\% of commits. This inverted pattern (higher file concentration, lower PR concentration) suggests bulk, coordinated changes rather than repeated debugging. Projects such as \texttt{airbytehq/airbyte}\footnote{\url{https://github.com/airbytehq/airbyte}}, with 810 unique YAML files changed across 109 PRs, illustrate sophisticated configuration management.

\begin{table}[t]
    \centering
    \caption{YAML Modification Rates by AI Agent}
    \vspace{-10pt}
    \label{tab:agent_yaml_rates}
    \begin{tabular}{p{2.8cm}rrr}
    \toprule
    \textbf{Agent} & \textbf{Total Files} & \textbf{YAML Files} & \textbf{YAML \%} \\
    \midrule
    Devin          & 169,467 & 8,186 & 4.83\% \\
    Cursor         & 37,838  & 1,661 & 4.39\% \\
    Claude Code    & 23,296  & 972   & 4.17\% \\
    Copilot        & 187,247 & 6,426 & 3.43\% \\
    OpenAI Codex   & 294,075 & 5,916 & 2.01\% \\
    \midrule
    \textbf{Total} & \textbf{711,923} & \textbf{23,161} & \textbf{3.25\%} \\
    \bottomrule
    \multicolumn{4}{l}{\small $\chi^2=3016.26$, $p<0.001$}
    \end{tabular}
    \vspace{-12pt}
\end{table}

\begin{figure}[ht]
    \centering
    \vspace{-8pt}
    \includegraphics[width=1\linewidth]{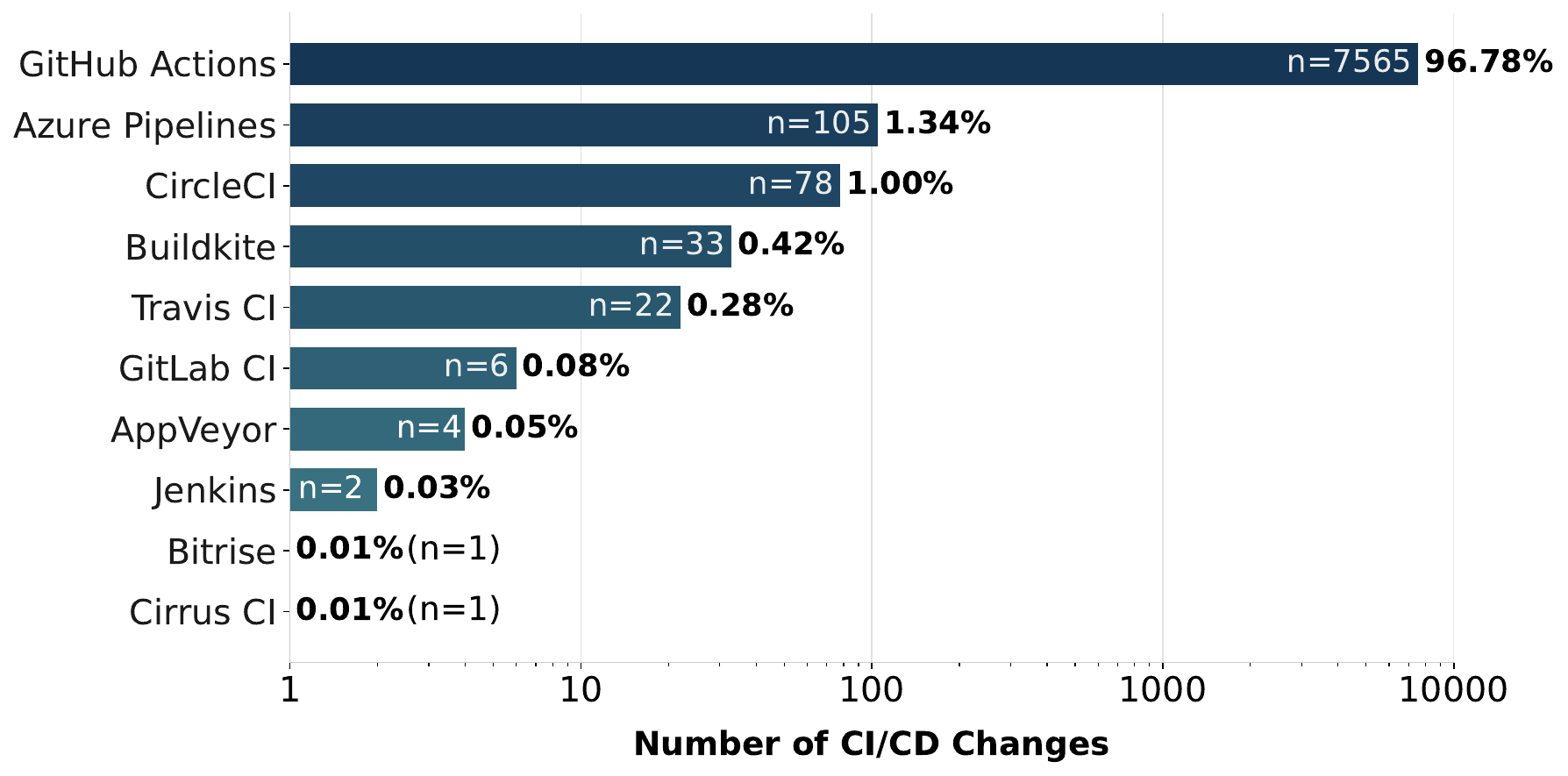}
    \vspace{-15pt}
    \caption{CI/CD Service Distribution (Top 10 of 24) -- the other 14 platforms had 0 CI/CD changes}
    \Description{CI/CD Service Distribution (Top 10 of 24). The other 14 platforms had no CI/CD changes.}
    \label{tab:ci_service_distribution}
    \vspace{-13pt}
\end{figure}

\smallskip
\noindent\textbf{\textsf{GitHub~Actions} accounts for 96.77\% of CI/CD changes.}
Of 23,161 YAML changes, 7,817 (33.75\%) target CI/CD platforms, so one-third of AI YAML changes concern CI/CD configuration. Figure~\ref{tab:ci_service_distribution} shows strong platform concentration: \textsf{GitHub~Actions} has 7,565 changes (96.77\%), \textsf{Azure~Pipelines} 105 (1.34\%), \textsf{CircleCI} 78 (1.00\%), \textsf{Travis~CI} 22 (0.28\%), and \textsf{GitLab~CI} 6 (0.08\%); 14 of 24 platforms received no changes.
Agents show significant CI/CD platform preferences ($\chi^2=435.38$, $p<0.001$). All favor \textsf{GitHub~Actions} (93.88–99.62\%), but corporate ownership introduces bias: Copilot performs 104 \textsf{Azure~Pipelines} changes (4.22\% of its CI/CD changes) versus 0–1 for other agents, a 42–84$x$ gap aligned with Microsoft ownership. OpenAI Codex is the most diverse, with 62 CircleCI changes (2.84\%) versus 2–6 for other agents, and is the only agent touching \textsf{Jenkins}, \textsf{Cirrus~CI}, and \textsf{GitLab~CI}. Devin and Cursor are almost exclusive to \textsf{GitHub~Actions} (99.62\% and 98.92\%).
Among 452 projects with at least five YAML changes, 173 (38.27\%) direct 75–100\% of YAML changes to CI/CD, confirming CI/CD as the primary context. We found only 20 projects (4.4\%) use multiple CI/CD platforms, indicating a strong single-platform tendency, likely due to deliberate simplification or limited multi-CI/CD agent support.

\subsection{RQ2: To what extent do CI/CD changes correlate with PR acceptance rates?}

\subsubsection{\textbf{Motivation.}}
Merge rates provide an indirect quality signal: if PRs changing CI/CD configurations are rejected more often than regular code changes, this may indicate (1) lower-quality configurations from agents, (2) stricter scrutiny of CI/CD due to its criticality, or (3) greater complexity and error-proneness of CI/CD pipeline changes. Examining whether this pattern is consistent across agents or varies by agent helps identify which ones may need additional training for configuration tasks. Merge rate differences can also reflect reviewer trust in agent-generated CI/CD configurations.

\subsubsection{\textbf{Approach.}}
We classified PRs by whether they modified CI/CD YAML. Using PR merge status, we computed: (1) overall merge rates for CI-changing versus non-CI/CD  PRs, (2) agent-specific merge rates for both categories, and (3) chi-square tests to determine statistical significance of differences. We constructed contingency tables crossing change type with merge outcome and performed tests on both aggregate data and per agent.

\subsubsection{\textbf{Findings.}}
Our analysis of 2,383 PRs with CI/CD YAML changes and 31,197 PRs without revealed that CI/CD configuration changes merge at slightly lower rates (67.77\%) compared to non-CI/CD  changes (71.80\%), a statistically significant 4.03pp difference ($\chi^2=12.95$, $p<0.0001$, odds ratio $=1.21$). However, this overall pattern masks substantial agent-level variation.

\smallskip
\noindent\textbf{CI/CD changes have slightly lower merge rates.}
Our analysis indicates that PRs changing CI/CD YAML have a 67.77\% merge rate (1,615 of 2,383) compared to 71.80\% for non-CI/CD changes (22,399 of 31,197). An odds ratio of 1.21 indicates PRs without CI/CD changes are 1.21$x$ more likely to be merged, suggesting slightly higher scrutiny for CI/CD changes but no fundamental issues. This difference is statistically significant ($p<0.0001$) but of limited practical importance, as most CI/CD changes are still merged.

\smallskip
\noindent\textbf{Copilot exhibits configuration specialization, with CI/CD changes being merged 15.63pp higher than application code.}
Table~\ref{tab:agent_merge_comparison} shows strong agent-level heterogeneity. Copilot shows an \textit{inverted} pattern: CI/CD YAML changes merge at 58.67\% versus 40.92\% for non-CI/CD changes, a 17.75pp advantage ($p<0.0001$). This suggests Copilot’s CI/CD changes are preferred over its general code, possibly because configuration changes receive closer review or because Copilot is stronger at configuration than general coding. This aligns with Copilot’s \textsf{Azure~Pipelines} performance in RQ1, indicating YAML specialization.
Conversely, OpenAI Codex shows strong, stable performance across both categories (77.01\% CI/CD versus 82.89\% non-CI/CD, -5.87pp, $p<0.0001$), with the highest merge rates for both change types. Devin has a modest but significant edge on CI/CD changes (63.37\% versus 52.52\%, +10.85pp, $p<0.0001$), while Cursor and Claude Code show no significant differences ($p=0.2225$ and $p=0.8701$), indicating comparable PR quality for CI/CD and non-CI/CD changes.

\begin{table}[ht]
    \centering
    \small
    \vspace{-4pt}
    \caption{Merge Rates by Agent and Change Type}
    \vspace{-9pt}
    \label{tab:agent_merge_comparison}
    \small
    \begin{tabular}{p{1.65cm}rrrrr}
    \toprule
    \textbf{Agent} & \textbf{CI/CD} & \textbf{Non-CI/CD} & \textbf{Diff (pp)} & \textbf{$\chi^2$} & \textbf{$p$} \\
    \midrule
    OpenAI Codex & 77.01\% & 82.89\% & -5.87  & 2.91   & <0.0001* \\
    Devin        & 63.37\% & 52.52\% & +10.85 & 23.58  & <0.0001* \\
    Claude Code  & 61.02\% & 58.90\% & +2.12  & 0.03   & 0.8701 \\
    Cursor       & 60.00\% & 65.74\% & -5.74  & 1.49   & 0.2225 \\
    Copilot      & 58.67\% & 40.92\% & +17.75 & 67.03  & <0.0001* \\
    \bottomrule
    \multicolumn{6}{l}{\footnotesize * $p < 0.05$ indicates statistically significant difference} \\
    \end{tabular}
    \vspace{-9pt}
\end{table}

\smallskip
\noindent\textbf{Absolute merge rate differences across agents dwarf the CI/CD modification effect.}
Differences in merge rate across agents are far larger than the CI/CD versus non-CI/CD effect. OpenAI Codex has 77.01--82.89\% merge rates across both categories, substantially outperforming all other agents. In contrast, Copilot's non-CI/CD  merge rate of 40.92\%, which is the lowest of all agent-category combinations, means fewer than half of its general code contributions are merged. The 41.97pp gap between OpenAI Codex's (82.89\%) and Copilot's (40.92\%) non-CI/CD performance is over 10 times larger than the 4.03pp overall CI/CD effect, indicating that agent selection matters far more than whether CI/CD is modified.

\subsection{RQ3: To what extent do CI/CD changes correlate with build success rates?}

\subsubsection{\textbf{Motivation.}}
While RQ2 examines whether CI/CD changes are \textit{merged} by human reviewers (merge rates), a critical question remains whether they \textit{work correctly} when executed. Build success rates provide direct measurement of CI/CD pipeline correctness. If agent-modified CI/CD configurations fail builds more frequently, this indicates agents struggle with configuration syntax, semantics, or integration requirements. This question is particularly important because CI/CD failures waste computational resources, delay development workflows, and erode developer trust in automation~\cite{zhao2017impact}.

\vspace{-1pt}
\subsubsection{\textbf{Approach.}}
We analyzed 99,930 completed \textsf{GitHub~Actions} workflow runs across 31,048 PRs. We classified each PR as changing or not changing \textsf{GitHub~Actions} YAML based on file change analysis. For each workflow run, we extracted execution outcome (\textit{success}, \textit{failure}, or \textit{canceled}) and computed: (1) overall success rates for PRs changing versus not changing \textsf{GitHub~Actions} configurations, (2) agent-specific success rates for both categories, and (3) chi-square tests to determine statistical significance of differences.

\vspace{-1pt}
\subsubsection{\textbf{Findings.}}
Our analysis revealed that PRs changing CI/CD YAML have nearly comparable success rates (75.59\%) to those that do not (74.87\%), as the difference is not statistically significant ($\chi^2=2.20$, $p=0.138$). This does not support the assumption that AI agents introduce reliability issues when changing configurations.

\smallskip
\noindent\textbf{Three of five agents show statistically significant improvements when changing CI/CD.}
Table~\ref{tab:ci_reliability_impact} shows that four agents (Claude Code, Copilot, Cursor, Devin) have \textit{higher} success rates when changing CI/CD YAML, with gains from 1.91pp (Devin) to 6.08pp (Claude Code). Three of these gains are statistically significant: Copilot (+5.98pp, $\chi^2=40.62$, $p<0.0001$), Cursor (+4.03pp, $\chi^2=4.80$, $p=0.029$), and Devin (+1.91pp, $\chi^2=5.89$, $p=0.015$). Claude Code’s 6.08pp gain is marginal ($p=0.051$), and OpenAI Codex shows virtually no difference (-0.12pp, $p=0.930$).
These results suggest that CI/CD configuration changes are either (a) crafted more carefully due to perceived complexity, (b) tested more thoroughly before PR submission, or (c) typically simpler changes (e.g., adding workflows) rather than complex refactorings.

\begin{table}[ht]
    \centering
    \vspace{-5pt}
    \caption{Build Success Rates of CI/CD YAML Changes}
    \vspace{-9pt}
    \label{tab:ci_reliability_impact}
    \small
    \begin{tabular}{p{2.18cm}rrrrr}
    \toprule
    \textbf{Agent} & \textbf{No CI} & \textbf{With CI} & \textbf{Diff (pp)} & \textbf{$\chi^2$} & \textbf{$p$} \\
    \midrule
    Claude Code  & 72.36\% & 78.44\% & +6.08 & 3.81  & 0.051 \\
    Copilot      & 68.77\% & 74.75\% & +5.98 & 40.62 & 0.000* \\
    Cursor       & 79.10\% & 83.13\% & +4.03 & 4.80  & 0.029* \\
    Devin        & 81.26\% & 83.16\% & +1.91 & 5.89  & 0.015* \\
    OpenAI Codex & 64.12\% & 64.00\% & -0.12 & 0.01  & 0.930 \\
    \midrule
    \textbf{Overall} & \textbf{74.87\%} & \textbf{75.59\%} & \textbf{+0.72} & \textbf{2.20} & \textbf{0.138} \\
    \bottomrule
    \multicolumn{6}{l}{\footnotesize * $p < 0.05$ indicates statistically significant difference} \\
    \end{tabular}
    \vspace{-9.3pt}
\end{table}

\smallskip
\noindent\textbf{Absolute success rates vary more by agent than by CI/CD modification.}  
Success rates differ far more by agent than by CI/CD changes. Devin performs best (81.26–83.16\%), followed by Cursor (79.10–83.13\%), with OpenAI Codex lowest (64.00–64.12\%), a 19–26pp gap, 26–36$\times$ larger than the 0.72pp overall effect of CI/CD changes, indicating that agent choice has much greater impact than changing CI/CD.  
Copilot’s 5.98pp gain with CI/CD changes ($p<0.0001$) mirrors its CI/CD merge-rate advantage in RQ2 (+15.63pp). This alignment between merge acceptance and build success suggests Copilot is genuinely stronger on CI/CD tasks than general code, supporting the YAML configuration specialization hypothesis.

\vspace{-2pt}
\section{Implications}
\label{sec:implications}
\vspace{-1pt}
\noindent\textbf{For AI Agent Developers.}
AI agent developers should address training data and engagement biases. The dominance of \textsf{GitHub~Actions} (96.77\%) creates a feedback loop that reinforces platform skew and hinders alternative adoption. Mitigations include incorporating diverse CI/CD platforms in training, disclosing platform biases, and evaluating performance across multiple CI/CD platforms. Specialized CI/CD training can further improve engagement and reliability by teaching agents when to change configurations, exposing them to best practices, and covering CI-specific failure modes.

\smallskip
\noindent\textbf{For Practitioners.}
Practitioners can leverage agent-specific strengths when deploying AI assistants. Devin’s high YAML engagement and strong build success suit frequent CI/CD updates, Codex provides consistent quality across platforms, and Copilot’s CI/CD specialization (+15.63pp merge rate, +5.98pp build success) favors infrastructure tasks over general coding. While most CI/CD changes may not need extra scrutiny, tools highlighting agent-specific risks and automating CI/CD validation can help manage heterogeneity, especially for specialized agents like Copilot.

\smallskip
\noindent\textbf{For Researchers.}
Researchers should adopt domain-specific evaluation and measure platform biases. In addition, Copilot’s paradox (i.e., higher CI/CD success than general code) shows that aggregate metrics hide specialized capabilities, demanding configuration-specific benchmarks for CI/CD and related infrastructure. Platform bias, such as \textsf{GitHub~Actions} concentration and Copilot’s \textsf{Azure~Pipelines} skew, requires systematic measurement and mitigation strategies. Cross-platform replication and context-aware assessment are essential to understand agent behavior, prevent overgeneralization, and guide research and policy on AI-assisted software configuration.

\vspace{-2pt}
\section{Threats to Validity}
\label{sec:threats}
\vspace{-1pt}

\noindent\textbf{Construct Validity.} Our CI/CD file detection relies on filename regular expression patterns, validated manually on random sample with nearly perfect precision, as misclassifications involved CI/CD platforms not under consideration or generally unknown YAML patterns. Moreover, quality proxies (i.e., merge rates and build success) may not fully capture correctness or code quality, but provide consistent large-scale signals, and agent-level comparisons are robust to systematic measurement errors.

\medskip
\noindent\textbf{Internal Validity.} Agent attribution uses heuristics (commit messages, PR descriptions, user declarations) with 94\% manual accuracy~\cite{li2025aidev}, likely causing conservative effect dilution. Temporal confounding may arise from agent capability evolution, and selection bias exists because only submitted PRs are analyzed, underrepresenting rejected or previewed suggestions, particularly for Copilot.

\smallskip
\noindent\textbf{External Validity.} All data comes from the AIDev dataset and GitHub, limiting generalization to other AI agents or PRs. \textsf{GitHub~Actions} dominance (96.77\%) likely reflects true agent bias but may differ elsewhere. The dataset emphasizes popular open-source repositories, thus results may not extend to other projects, specialized CI/CD needs, or private enterprise projects with different workflows and acceptance practices.

\section{Related Work}
\label{sec:related}

\vspace{-1pt}
\noindent\textbf{AI-Assisted Code Generation.} Nguyen et al.~\cite{nguyen2022empirical} found GitHub Copilot's suggestions vary in quality with task complexity. Perry et al.~\cite{perry2023users} showed developers using AI assistants write more insecure code, and Asare et al.~\cite{asare2023github} found Copilot introduces vulnerabilities at rates comparable to humans. Recent studies applied LLMs to generate~\cite{ghaleb2025llm4ci} and migrate~\cite{hossain2025cigrate} CI configurations, but they do not characterize AI agent behavior in configuration modification.

\smallskip
\noindent\textbf{Continuous Integration Practices.} Hilton et al.~\cite{hilton2016usage} analyzed 34,544 open-source projects, finding 40\% use CI/CD. Zhao et al.~\cite{zhao2017impact} demonstrated CI/CD adoption improves code review speed but requires non-trivial configuration effort. Vasilescu et al.~\cite{vasilescu2015quality} showed CI/CD usage correlates with higher merge rates. Recent work on CI/CD adoption~\cite{chopra2025multici,rostami2023usage} and configuration practices~\cite{ghaleb2025android,abrokwah2025empirical,delicheh2026automation} analyzed the complexity and maintenance challenges of these systems, but did not analyze how AI agents could contribute in addressing them.

\smallskip
\noindent\textbf{Pull Request Quality.} Gousios et al.~\cite{gousios2014exploratory} characterized PR practices, identifying factors influencing merge decisions. Tsay et al.~\cite{tsay2014influence} showed that social and technical factors jointly predict PR acceptance. Our work examines whether agent-generated PRs receive different treatment when changing CI configurations.

\vspace{-1pt}
\section{Conclusion}
\label{sec:conclusion}
\vspace{-1pt}
We present the first empirical study of how AI agents interact with CI/CD configurations, analyzing 711,923 file changes across 8,031 PRs from 1,605 GitHub repositories. We find: (1) limited engagement with CI/CD (3.25\%), with significant agent variation ($\chi^2=3016$, $p<0.001$); (2) extreme \textsf{GitHub~Actions} dominance (96.77\%) and corporate effects (Copilot's 42-84$x$ Azure bias) threaten platform diversity; (3) agents handle changes reliably, with three showing improved build success. Notably, Copilot exhibits a configuration specialization paradox, merging CI/CD changes 15.63pp more often and achieving 5.98pp higher build success than general code. These results indicate developers should address training data bias, develop CI-specific agent capabilities, and adopt evidence-based agent selection, reserving Copilot for CI/CD tasks where it excels.

\smallskip
\noindent\textbf{Future work} should explore controlled experiments on Copilot’s CI/CD advantage, longitudinal studies of agent evolution, multi-platform replications, and configuration-specialized agents. As AI becomes central to software development, continued empirical scrutiny of agent interactions with CI/CD is essential, and our study offers a baseline for future comparisons.

\vspace{-1pt}
\begin{acks}
\vspace{-1pt}
This work is funded by the Natural Sciences and Engineering Research Council of Canada (NSERC): RGPIN-2025-05897.
\end{acks}

\clearpage
\balance
\bibliographystyle{ACM-Reference-Format}
\bibliography{refs}

@article{asare2023github,
  title={Is {GitHub's Copilot} as bad as humans at introducing vulnerabilities in code?},
  author={Asare, Owura and Nagappan, Meiyappan and Asokan, Nirmal},
  journal={Empirical Software Engineering},
  volume={28},
  number={6},
  pages={129},
  year={2023},
  publisher={Springer}
}

@article{fitzgerald2017continuous,
  title={Continuous software engineering: A roadmap and agenda},
  author={Fitzgerald, Brian and Stol, Klaas-Jan},
  journal={Journal of Systems and Software},
  volume={123},
  pages={176--189},
  year={2017},
  publisher={Elsevier}
}

@article{li2025aidev,
  title={{The Rise of AI Teammates in Software Engineering (SE) 3.0: How Autonomous Coding Agents Are Reshaping Software Engineering}}, 
  author={Li, Hao and Zhang, Haoxiang and Hassan, Ahmed E.},
  journal={arXiv preprint arXiv:2507.15003},
  year={2025}
}

@article{shahin2017continuous,
  title={Continuous integration, delivery and deployment: a systematic review on approaches, tools, challenges and practices},
  author={Shahin, Mojtaba and Babar, Muhammad Ali and Zhu, Liming},
  journal={IEEE access},
  volume={5},
  pages={3909--3943},
  year={2017},
  publisher={IEEE}
}

@article{watanabe2025use,
  title={On the use of agentic coding: An empirical study of pull requests on {GitHub}},
  author={Watanabe, Miku and Li, Hao and Kashiwa, Yutaro and Reid, Brittany and Iida, Hajimu and Hassan, Ahmed E},
  journal={arXiv preprint arXiv:2509.14745},
  year={2025}
}

@inproceedings{gousios2014exploratory,
  title={An exploratory study of the pull-based software development model},
  author={Gousios, Georgios and Pinzger, Martin and Deursen, Arie van},
  booktitle={Proceedings of the 36th International Conference on Software Engineering},
  pages={345--355},
  year={2014}
}

@inproceedings{hilton2016usage,
  title={Usage, costs, and benefits of continuous integration in open-source projects},
  author={Hilton, Michael and Tunnell, Timothy and Huang, Kai and Marinov, Darko and Dig, Danny},
  booktitle={Proceedings of the 31st IEEE/ACM International Conference on Automated Software Engineering},
  pages={426--437},
  year={2016}
}

@inproceedings{nguyen2022empirical,
  title={An empirical evaluation of {GitHub Copilot's} code suggestions},
  author={Nguyen, Nhan and Nadi, Sarah},
  booktitle={Proceedings of the 19th International Conference on Mining Software Repositories},
  pages={1--5},
  year={2022}
}

@inproceedings{perry2023users,
  title={Do users write more insecure code with {AI} assistants?},
  author={Perry, Neil and Srivastava, Megha and Kumar, Deepak and Boneh, Dan},
  booktitle={Proceedings of the 2023 ACM SIGSAC conference on computer and communications security},
  pages={2785--2799},
  year={2023}
}

@inproceedings{sandoval2023lost,
  title = {Lost at {C}: A User Study on the Security Implications of Large Language Model Code Assistants},
  author = {Sandoval, Gustavo and Pearce, Hammond and Tan, Teo and Karri, Ramesh and Dolan-Gavitt, Brendan},
  booktitle = {32nd USENIX Security Symposium},
  year = {2023},
  pages = {2205--2222},
}

@inproceedings{tsay2014influence,
  title={Influence of social and technical factors for evaluating contribution in {GitHub}},
  author={Tsay, Jason and Dabbish, Laura and Herbsleb, James},
  booktitle={Proceedings of the 36th International Conference on Software Engineering},
  pages={356--366},
  year={2014}
}

@inproceedings{vasilescu2015quality,
  title={Quality and productivity outcomes relating to continuous integration in {GitHub}},
  author={Vasilescu, Bogdan and Yu, Yue and Wang, Huaimin and Devanbu, Premkumar and Filkov, Vladimir},
  booktitle={Proceedings of the 2015 10th joint meeting on foundations of software engineering},
  pages={805--816},
  year={2015}
}

@inproceedings{zhao2017impact,
  title={The impact of continuous integration on other software development practices: a large-scale empirical study},
  author={Zhao, Yangyang and Serebrenik, Alexander and Zhou, Yuming and Filkov, Vladimir and Vasilescu, Bogdan},
  booktitle={2017 32nd IEEE/ACM International Conference on Automated Software Engineering (ASE)},
  pages={60--71},
  year={2017},
  organization={IEEE}
}

@misc{replication_package,
  title={When AI Agents Touch CI/CD Configurations: Frequency and Success (Replication Package)},
  author={Taher A. Ghaleb},
  howpublished={\url{https://github.com/Taher-Ghaleb/AIAgentsCI_CD-MSR2026}},
  year={2026}
}

@article{fowler2024continuous,
  title={{Continuous integration}},
  year={2024},
  author={Fowler, Martin},
  journal={\url{https://martinfowler.com/articles/continuousIntegration.html}}
}

@article{mchugh2013chi,
  title={The {Chi-square} test of independence},
  author={McHugh, Mary L},
  journal={Biochemia medica},
  volume={23},
  number={2},
  pages={143--149},
  year={2013},
  publisher={Medicinska naklada}
}

@article{ghaleb2019empirical,
  title={An empirical study of the long duration of continuous integration builds},
  author={Ghaleb, Taher Ahmed and Da Costa, Daniel Alencar and Zou, Ying},
  journal={Empirical Software Engineering},
  volume={24},
  pages={2102--2139},
  year={2019},
  publisher={Springer}
}

@article{ghaleb2019noise,
  title={Studying the impact of noises in build breakage data},
  author={Ghaleb, Taher Ahmed and Da Costa, Daniel Alencar and Zou, Ying and Hassan, Ahmed E},
  journal={IEEE Transactions on Software Engineering},
  volume={47},
  number={9},
  pages={1998--2011},
  year={2019},
  publisher={IEEE}
}

@article{ghaleb2022interplay,
  title={Studying the interplay between the durations and breakages of continuous integration builds},
  author={Ghaleb, Taher A and Hassan, Safwat and Zou, Ying},
  journal={IEEE Transactions on Software Engineering},
  volume={49},
  number={4},
  pages={2476--2497},
  year={2022},
  publisher={IEEE}
}

@article{ghaleb2025android,
  title={CI/CD Configuration Practices in Open-Source Android Apps: An Empirical Study},
  author={Ghaleb, Taher and Abduljalil, Osamah and Hassan, Safwat},
  journal={ACM Transactions on Software Engineering and Methodology},
  year={2024},
  publisher={ACM New York, NY}
}

@inproceedings{ghaleb2025llm4ci,
  title={Can {LLMs} Write {CI}? A Study on Automatic Generation of GitHub Actions Configurations},
  author={Ghaleb, Taher A and Rathnayake, Dulina},
  booktitle={2025 IEEE International Conference on Software Maintenance and Evolution (ICSME)},
  pages={767--772},
  year={2025},
  organization={IEEE}
}

@inproceedings{chopra2025multici,
  title={From First Use to Final Commit: Studying the Evolution of Multi-CI Service Adoption},
  author={Chopra, Nitika and Ghaleb, Taher A},
  booktitle={2025 IEEE International Conference on Software Maintenance and Evolution (ICSME)},
  pages={773--778},
  year={2025},
  organization={IEEE}
}

@article{hossain2025cigrate,
  title={{CIgrate}: Automating {CI} Service Migration with Large Language Models},
  author={Hossain, Md Nazmul and Ghaleb, Taher A},
  journal={arXiv preprint arXiv:2507.20402},
  year={2025}
}

@article{abrokwah2025empirical,
  title={An Empirical Study of Complexity, Heterogeneity, and Compliance of {GitHub Actions} Workflows},
  author={Abrokwah, Edward and Ghaleb, Taher A},
  journal={arXiv preprint arXiv:2507.18062},
  year={2025}
}

@article{rostami2023usage,
  title={On the usage, co-usage and migration of CI/CD tools: A qualitative analysis},
  author={Rostami Mazrae, Pooya and Mens, Tom and Golzadeh, Mehdi and Decan, Alexandre},
  journal={Empirical Software Engineering},
  volume={28},
  number={2},
  pages={52},
  year={2023},
  publisher={Springer}
}

@article{zheng2025github,
  title={Why Do GitHub Actions Workflows Fail? An Empirical Study},
  author={Zheng, Lianyu and Li, Shuang and Huang, Xi and Huang, Jiangnan and Lin, Bin and Chen, Jinfu and Xuan, Jifeng},
  journal={ACM Transactions on Software Engineering and Methodology},
  year={2025},
  publisher={ACM New York, NY}
}

@article{delicheh2026automation,
  title={Automation and Reuse Practices in GitHub Actions Workflows: A Practitioner's Perspective},
  author={Delicheh, Hassan Onsori and Cardoen, Guillaume and Decan, Alexandre and Mens, Tom},
  journal={arXiv preprint arXiv:2601.11299},
  year={2026}
}

\end{document}